\documentclass[twocolumn,english,floatfix,aps,prb]{revtex4}
\usepackage{graphicx}
\usepackage{amssymb}

\makeatletter

\usepackage{slashbox}
\usepackage{hhline}
\usepackage{psfrag}

\usepackage{babel}
\makeatother
\begin{document}

\title{Matching Kasteleyn Cities for Spin Glass Ground States}

\author{Creighton K. Thomas, A. Alan Middleton}

\affiliation{Department of Physics, Syracuse University, Syracuse, NY 13244, USA}

\begin{abstract}
As spin glass materials have extremely slow dynamics, devious numerical
methods are needed to study low-temperature states. A simple and fast
optimization version of the classical Kasteleyn treatment of the Ising
model is described and applied to two-dimensional Ising spin glasses.
The algorithm combines the Pfaffian and matching approaches to directly
strip droplet excitations from an excited state. Extended ground states
in Ising spin glasses on a torus, which are optimized over all boundary
conditions, are used to compute precise values for ground state energy
densities.
\end{abstract}
\maketitle
The Ising spin glass is a model for disordered magnetic alloys which
captures the complexity of materials with frozen randomness and competing
interactions, including frustration, extremely slow dynamics, and
intricate memory effects.\cite{YoungBook} The spins in the model
are coupled by random choices of ferromagnetic or antiferromagnetic
bonds, leading to a complex free energy landscape. There are at
least two theoretical approaches,\cite{pictures} including the droplet
and replica-symmetry-breaking pictures,
used to describe the non-equilibrium dynamics and low-free-energy
structure of the spin glass phase space. As these theoretical approaches
differ significantly and exact results for spin glasses are
rare,\cite{Exactish}
computational work has been essential for computing
scaling exponents and as a qualitative check of theoretical
pictures.\cite{OptBook}

The history of the relationship between the physical analysis and
the mathematics of numerical approaches to spin glasses is long and
rich.
In general, characterizing the complex free energy landscape
of disordered materials is challenging.
Direct Monte Carlo simulations are hindered
by the same high free-energy barriers that inhibit equilibration in
the physical system. It is expected\cite{YoungBook,pictures} that
times $t$ satisfying $\ln(t)\sim L^{\psi}$ are required to equilibrate
systems of size $L$, where $\psi\ge\theta$ and $\theta$ determines
the energy scale $\Delta E(\ell)$ of excitations or domain walls
on length scale $\ell$, $\Delta E\sim\ell^{\theta}$. To replicate
the many decades of experimental time scales and to develop a better
understanding of disordered systems for $t\rightarrow\infty$, algorithms
for either accelerating the approach to equilibrium or finding ground states
in spin glasses have been developed.
Many of these techniques
(which are often generally applicable to
disordered materials)
are inspired by, or have inspired,
approaches for combinatorial optimization problems.
Parallel tempering, genetic algorithms,
and extremal optimization are examples of heuristic algorithms to
find close approximations to equilibrated and ground state
configurations.\cite{Heuristics}
Exact general algorithms such as transfer matrix
methods\cite{Transfer} and branch-and-cut methods\cite{BranchCut}
require times that are exponential in powers of the system size,
though extensive development has led to computing ground states in
three-dimensional Ising spin glasses with up to $12^{3}$ spins.\cite{BranchCut}

We have found a simple algorithm for studying two-dimensional (2D)
Ising spin glasses that combines use of the classical Kasteleyn
city\cite{Kasteleyn} and application of a standard combinatorial optimization
algorithm. Besides solving the problem on planar graphs and linking
together these methods, we use this algorithm to study ``extended''
ground states, which optimize the energy over choices of periodic
or anti-periodic boundary conditions, as well as the spin configuration.
This approach dramatically improves the treatment of boundary-free
samples, so that the finite-size effects are greatly reduced. We have
used this algorithm to determine very precisely the energy of the
Ising spin glass in the large volume limit.

The Edwards-Anderson Hamiltonian that is used for Ising spin glasses
is ${\cal H}\left(\left\{ s_{i}\right\} \right)=-\sum_{\langle ij\rangle}J_{ij}s_{i}s_{j}$,
where the couplings $J_{ij}$ between nearest neighbor pairs of spins
$\langle ij\rangle$ are independent identically distributed variables,
fixed in a given sample, and the $s_{i}$ are Ising spin variables,
$s_{i}=\pm1$, with $L^{d}$ sites $i$ on a $d$-dimensional lattice.
The distribution for $J_{ij}$ is generally taken either to be Gaussian
or bimodal (the ``$\pm J$'' case). Barahona\cite{Barahona}
has shown that computing the ground state energy of a 3D spin glass
(or even two coupled 2D layers) is an NP-hard problem.\cite{TSP}
This implies that if the ground state of the 3D spin glass could be
efficiently computed, i.e., found in a time polynomial in $L$, many outstanding
computational problems that are believed to require worst-case
exponential time
to solve, such as the Traveling Salesperson Problem, could also be
solved in time polynomial in the size of the problem. Improvements
in 3D spin glass calculations therefore focus on reducing the numerical
constants in the exponent for the expected solution time. 

The two-dimensional Ising spin glass (2DISG) is a case where exact
algorithms have allowed for study of the ground state and density
of states. These approaches have used two methods: the dimer-Pfaffian
method (Pfaffian method) and matching to minimize frustration.

The partition function for Ising models with arbitrary couplings can
be solved for either open or toroidal boundary conditions by using
techniques developed for the pure Ising model,\cite{Kasteleyn} i.e.,
computing and summing Pfaffians, antisymmetric combinations of ordered
statistical
weights, from $L^{2}\times L^{2}$ sparse matrices.
The ground state energy can be computed\cite{PfaffianSpinGlass}
in $O(L^{5})$ time for discrete-valued disorder; the spectrum is
discrete and bounded by a power of $L$. Note that the Pfaffian method
uses perfect matchings (dimer coverings) on a decorated lattice
and requires a sum over four combinations of odd and even constraints
on these matchings on a torus.

\begin{figure}
\includegraphics[width=1\columnwidth]{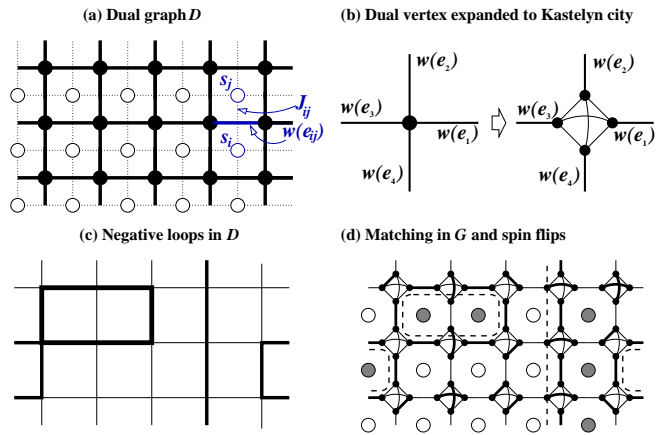}

\caption{\label{cap:GadgetGraph}(color online) An outline of
the steps that convert a spin and bond configuration first
to the dual weighted lattice $D$ and then to the weighted
graph $G$, in order to compute the extended ground state.
(a) The original spin system, here with initial spins
$s_{i}=1$ indicated by white circles, and bond configuration
$J_{ij}$ (dashed lines) determine edge weights
$w_{ij}=J_{ij}s_{i}s_{j}$ (taking the initial BCs to be
periodic) in the dual graph $D$
(solid vertices and edges), with periodic boundary
conditions. (b) The vertices in $D$ are replaced by
Kasteleyn cities (light lines have zero weight in $G$). (c)
An example set $S$ of negative weight loops in $D$ is shown,
with two simple loops and one winding loop. (d) Heavy lines
indicate the minimum weight perfect matching $M$ for $G$
(light lines are free edges and solid circles are vertices
in $G$). The negative weight loops $S$ (heavy dashed lines)
are found by clipping out the Kasteleyn cities and keeping
the remaining edges. Finally, spins are assigned by scanning
across the sample: each time an odd number of loops is
crossed, the spins are flipped (gray circles indicate
$s_{i}=-1$). In this case, the inconsistency at the right
side is corrected by changing horizontal boundary conditions
from periodic to antiperiodic.} \end{figure}

The fastest ground state algorithms for the 2DISG map
the Ising spin glass problem to the weighted perfect matching problem,
a common problem in combinatorial optimization. Given a graph $G=(V,E)$,
with vertices $V$ and edges $E$, with a weight function $w:E\rightarrow\mathbb{R}$,
the problem is to select a perfect matching, a subset of edges $M\subset E$
where every vertex in $V$ belongs to a single edge $e\in E$, such
that the total weight $w(M)=\sum_{e\in M}w_{e}$ is minimal. The solution
can be found in time polynomial in the number of edges.\cite{TSP}
Matching is the core routine in two mappings for finding 2DISG ground
states. The mapping by Bieche \emph{et al}.\cite{Bieche}\ uses a
graph where the vertex set $V$ contains the frustrated plaquettes
(primitive polygons $p$ with $\Pi_{\langle ij\rangle\in p}J_{ij}<0$).
The edges connect points in $V$ within some
distance $r_{\mathrm{max}}$. This algorithm is
exact as $r_{\mathrm{max}}\rightarrow\infty$, but it works
for a large fraction of cases even with small values of
$r_{\mathrm{max}}$, especially for $\pm J$ disorder.
Barahona's mapping\cite{Barahona}
replaces each plaquette with a subgraph that is connected to neighboring
subgraphs by dual bonds, with each dual bond crossing one edge in
G [see Fig. \ref{cap:GadgetGraph}(a)].
The subgraph edges have zero weight and the
dual edges that cross bonds of strength $J_{ij}$
have weight $w_{ij}=\left|J_{ij}\right|$; the subgraph comes in two types,
assigned according to the frustration of the plaquette. These algorithms
have been extremely useful, e.g., in studying domain walls and the
nature of the ground state as $L\rightarrow\infty$.\cite{CHK2004,SLEDW}
Note that the graphs used are derived from the (sample-dependent) plaquette frustrations;
this is not the case with our algorithm, where the graph is
independent of the $J_{ij}$, so its implementation is simpler.

Matching algorithms have been used for planar graphs. The case of
the torus, with periodic boundary conditions in both directions, has
not been addressed in very large systems, as Pfaffian methods are
much slower (and in practice, mean-time exponential run-time
algorithms are still commonly used).
Studies of smaller toroidal systems with Gaussian disorder
have used the branch-and-cut algorithm\cite{CHK2004} or the transfer
matrix; such studies confirm that the finite-size corrections vanish
much more quickly in toroidal geometries rather than planar geometries.
It would therefore be useful to have a fast algorithm for finding
information about the ground states for the 2DISG on the torus.

We have developed an approach which is not limited to planar
graphs; it also provides significant information about the
ground state on the torus. One component of this approach is
a ground-state algorithm that combines a representation from
the Pfaffian method with matching. The other component is
applying this algorithm on the torus to find an extended
ground state: the minimum energy state over all spin
configurations and over the set of four boundary conditions
(BCs). That is, we find the extended state $\left(\left\{
s_{i}^{0}\right\} ,\sigma_{h}^{0},\sigma_{v}^{0}\right)$
which minimizes ${\cal H}^*=-\sum_{\langle
ij\rangle}J_{ij}s_{i}s_{j}\sigma_{ij}$, with $\sigma_{ij}=1$
except on one vertical column of horizontal bonds, where
$\sigma_{ij}=\sigma_{v}$, and on one horizontal row of
vertical bonds, where $\sigma_{ij}=\sigma_{h}$ and
$\sigma_{h}$ and $\sigma_{v}$ take values
$\sigma_{h,v}=\pm1$. The extended ground state on the torus
is the minimum energy state over the four possible
combinations of BCs given by choosing (anti-)periodic BCs
for each direction.  The standard ground state for given BCs
is therefore exactly found for $\frac{1}{4}$ of the samples.  Note
that, in general, when all $\sigma_{ij} = 1$, ${\mathcal
H}^* = {\mathcal H}$, so the extended ground state is equal
to the standard ground state (this is always the case for
planar graphs, so the algorithm finds ground states
of planar graphs without modification).  The extended ground
state on the torus is also of interest in its own right.  For example, it
can be used as an edge-free background for studying
equilibration and droplets\cite{CMWprep} and to rapidly
compute the energy density for large samples.

We first give an overview of our algorithm.
A spin and bond configuration is used
to define a weighted dual lattice $D$ which in turn is mapped to
a weighted graph $G$. A minimum weight perfect matching for $G$ is computed and
used to identify a set of negative weight loops in $D$
with the most negative total weight. These loops are
exactly the excitations of the current configuration
relative to an extended ground state. The configuration is
thus set to
the ground state by flipping the spins ``within'' each loop.
This method can be applied to any planar graph by supplying
the appropriate boundary conditions (i.e.\ in the same way as
Bieche {\em et al}.\ and Barahona algorithms).

A more detailed description of the method
for the $L\times L$ toroidal square
lattice starts with a list of the inputs:
an initial configuration $c=(\{s_{i}\},\sigma_h,\sigma_v)$ and
couplings $J_{ij}$. The dual lattice
$D=(V,E)$ has edges $e_{ij}\in E$ ($e_{ij}$ crosses the bond
$\langle ij \rangle$ in the original lattice) connecting neighboring
plaquettes (these make up $V$) on the original spin lattice;
it also is an $L\times L$
torus. Given $c$, weights $w_{c}$ for edges in $E$
are set by $w_c(e_{ij})=J_{ij}s_{i}s_{j}\sigma_{ij}$;
see Fig.~\ref{cap:GadgetGraph}(a).
The value of the extended
Hamiltonian is then ${\cal
H}^*(c)=-w_{c}(E)\equiv-\sum_{e_{ij}\in E}w_{c}(e_{ij})$.

To minimize ${\cal H}^*$, we find the extremal (i.e.,
minimum total weight) set of negative weight loops in the
dual graph $D$ by computing a minimal weight perfect
matching $M$ on a related graph $G$. In the case of a square
lattice, we form $G$ by replacing each vertex in $D$ by a
``Kasteleyn city'' subgraph, a complete graph with $4$ nodes
[see Fig.~\ref{cap:GadgetGraph}(b)]; such mappings exist for
any lattice. Weights for edges in $G$ are zero on city edges
and are given by $w_{c}(e_{ij})$ on edges $e_{ij}$ kept from
$D$ (cf.\ the Barahona algorithm, which instead uses
$|w(e_{ij})|$, which is independent of $c$; frustration is
incorporated via the use of two distinct graph decorations).
Matchings in $G$ can be mapped to sets of loops in $D$: one
simply contracts out the Kasteleyn cities from $M$ to arrive
at a set of loops made of edges $S\subseteq E$ [see
Fig.~\ref{cap:GadgetGraph}(c,d)].  The Kasteleyn cities
enforce the constraint that an even number of edges in $S$
meet at each dual vertex (i.e.\ $S$ is a collection of
Eulerian subgraphs of $D$).

To prove the correctness of the algorithm, we first show
that the weight of the loops that relate two configurations
is proportional to the energy difference between the
configurations.  For an extended spin configuration $c$, let
$b_{ij}(c) = s_i s_j \sigma_{ij}$.  When comparing two
extended configurations $c$ and $c'$, call $S$ the set of
bonds in which $b_{ij}(c) = -b_{ij}(c')$ (note that
$b_{ij}(c) = \pm b_{ij}(c')$ always).  Since in $S$,
$b_{ij}(c) = -b_{ij}(c')$ and in $E\backslash S$, $b_{ij}(c)
= b_{ij}(c')$, we have that
\begin{eqnarray}
{\mathcal H}^*(c)-{\mathcal H}^*(c') & = &
                        -\sum_{\langle ij \rangle} J_{ij}b_{ij}(c)
			+\sum_{\langle ij \rangle}
			J_{ij}b_{ij}(c') \nonumber \\
                    &=& -\sum_{e_{ij}\in S} J_{ij}b_{ij}(c)
		        -\sum_{e_{ij}\in E\backslash S}
			J_{ij}b_{ij}(c) \nonumber \\
			& &
			+\sum_{e_{ij}\in S} J_{ij}b_{ij}(c')
		        +\sum_{e_{ij}\in E\backslash S}
			J_{ij}b_{ij}(c') \nonumber\\
	            &=& -2\sum_{e_{ij}\in S} J_{ij}b_{ij}(c)
		    \nonumber \\
	            &=& -2\sum_{e_{ij}\in S} w(e_{ij}(c)),
\end{eqnarray}
so that the energy difference between configurations is given by twice the
weight of $S$.

The proof that the minimum weight even-degree subgraph
always finds the ground state, then, is as follows.  Assume,
for the sake of contradiction, that there exists some
extended spin configuration $c^{0}$ with a lower total
energy than the $c'$ returned by our algorithm from initial
configuration $c$.  Call $S$ the set of bonds for which
$b_{ij}(c) = -b_{ij}(c')$, and $S^{0}$ the set of bonds for
which $b_{ij}(c) = -b_{ij}(c^{0})$.  Since ${\mathcal
H}^*(c^{0}) < {\mathcal H}^*(c')$, the energy difference
${\mathcal H}^*(c)- {\mathcal H}^*(c^{0}) > {\mathcal
H}^*(c) - {\mathcal H}^*(c')$ gives $2\sum_{e_{ij}\in
S^{0}}w_{c}(e_{ij}) < 2\sum_{e_{ij}\in S}w_c(e_{ij})$, which
means $S^{0}$ is an even-degree subgraph of D with a more
negative weight than $S$, in contradiction with the
assumption that $S$ is the extremal weight even degree
subgraph of $D$.

Note that Kasteleyn cities are often
described on the original lattice, where loops represent a high-temperature
expansion, but here on the dual lattice these loops contain clusters
in a low-temperature expansion.
The terms that contribute to the Pfaffian\cite{Kasteleyn} are products of statistical
weights $\pm e^{-\beta J_{ij}}$ over edges in loops in $D$ and statistical weights of unit norm
from the Kasteleyn cities. The dominant term in the Pfaffian that maximizes
the norm of such a product minimizes the sums of the $w_{ij}$ consistent
with a perfect matching in the graph $G$.
We note that there has been at least
one mention of using matching on the torus,\cite{LandryCoppersmith}
where one of the four ground states was found using the Bieche {\em et
al.} algorithm, but the utility of the extended ground state has been
made apparent and proven by this algorithmic framework.

This algorithm is simple to implement (given a standard
matching algorithm) and fast. On a 3.2 GHz Pentium IV
processor, the extended ground state for a $100^{2}$
square lattice on a torus is computed in $0.8\,\mathrm{s}$
for Gaussian disorder, where we use Blossom IV \cite{Blossom} for the matching routine. The mean
solution time scales approximately as $L^{3.5}$ through
toroidal lattices of size $400^{2}$. On toroidal graphs with $L\le 128$,
our algorithm, which finds exact ground states, is at least
three times as fast as our implementation of the Bieche
{\em et al}.\ algorithm. Note that the Bieche {\em et al}.\ algorithm does not
find the exact optimal state in in all cases -- in this case 
1.5\% of the samples (when $r_{\mathrm{max}}=8$).  Because the structure of the 
graph used in the Barahona algorithm is similar in structure
to that of our algorithm, the two algorithms have similar
performance, with
the Barahona algorithm using slightly less time (about 20\%)
and more memory (about 20\%).

On a torus, we use this algorithm to exactly 
solve for the extended ground state, which
is closely related to, but different than, finding the ground
state of the spin glass for given BCs.
The ground state energies for the four possible BCs differ by $O(L^{\theta})$,
which is the energy of a system-spanning domain wall.
The extended ground state, the minimum of the four, therefore has
at most an energy difference of $O(L^{\theta})$ from that for specified
BCs.

\begin{figure}
\includegraphics[width=1\columnwidth]{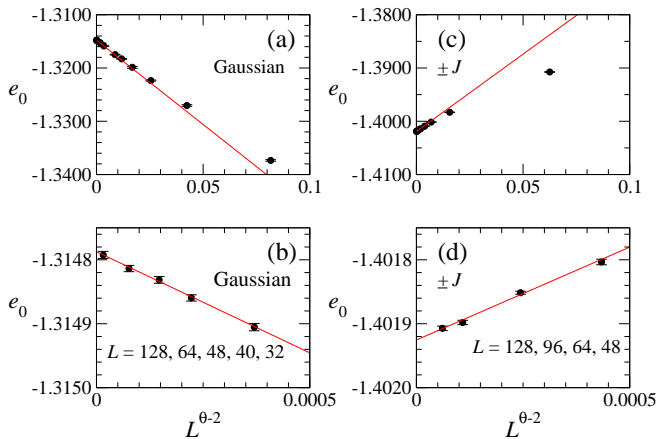}

\caption{\label{cap:Results}(color online) The extended ground state energy density $e_{0}(L)$
for the 2D Ising spin glass on a torus is plotted vs.~scaled system
size $L^{\theta-2}$. Two scales for each disorder type are used,
to show the linear fit at large $L$ and the higher-order corrections
at small $L$. (a,b) Assuming $\theta\approx-0.28$ for Gaussian disorder
gives $e_{0}(\infty)=-1.314788(4)$. (c,d) A similar plot using $\theta=0$
for discrete values of $J_{ij}=\pm1$ gives $e_{0}^{\pm}(\infty)=-1.401925(3)$.}
\end{figure}

This $O(L^{\theta})$ difference 
is the same order as the expected finite-size correction
to the ground state energy in a periodic system, so the extended ground
state is useful for studying energy densities.
We computed the sample
average of the extended ground state energy ${\cal H}^*_{0}$, using
at least $5\times10^{6}$ samples for $L\le64$ and at least $10^{6}$
samples for sizes $128\ge L>64$, both for Gaussian disorder ($\overline{J_{ij}^{2}}$=1,
$\overline{J_{ij}}=0$) and for the $\pm J$-model, $J_{ij}=\pm1$
with equal probability. We then plotted the sample average of the
ground-state energy density, $e_{0}(L)=\overline{{\cal H}^*_{0}L^{-2}}$,
vs\@.~$L^{\theta-2}$, which will give a straight line where
the leading finite-size correction dominates. For Gaussian disorder,
we find a linear fit to be very good for $L\ge32$, as shown in Fig.~\ref{cap:Results}(a,c),
for a wide range of $\theta\approx-0.28(4)$ ($\theta$ is not precisely
determined by this method; see a summary of results in
Ref.~\onlinecite{CHK2004})
and a highly precise estimate $e_{0}=-1.314788(4)$ (cf., e.g.,
$e_{0}=-1.31479(2)$ from Ref.~\onlinecite{CHK2004}). Taking $\theta=0$
for the $\pm J$ data also gives a good fit, with $e_{0}^{\pm}=-1.401925(3)$
(cf.\ $e_{0}^{\pm}=-1.40193(2)$ from Ref.~\onlinecite{PalmerAdler};
finite-size effects in our $L=48$ samples are less than those for $L=1800$
samples with open BCs). The extra precision results from the rapid
convergence to the thermodynamic limit in boundary-free samples, which
can be solved much faster than standard periodic samples solved using
branch-and-cut.


%

In conclusion, we have linked together Pfaffian and matching methods
to develop a fast algorithm for finding extended ground states in
the two-dimensional Ising spin glass on a torus or standard ground
states on planar graphs. For many purposes, the extended ground states
on a torus are as useful as ground states that are computed for a
fixed choice of periodic and antiperiodic boundary conditions, as
we show by precisely computing ground state energy densities. In the
Pfaffian method for computing the partition function $Z$ using the
dual lattice (i.e., low temperature expansion), the dominant term
in any of the four Pfaffians used to compute $Z$ is due to this extended
ground state; the partition function for a specified BC combination
is found by carefully cancelling out configurations with other boundary
conditions in the sum. Our method therefore is a combinatorial method,
based on matching, for finding the term that dominates the contributing
Pfaffians at low temperature. 

This work was supported in part by NSF grant DMR 0606424; we thank
the KITP (NSF grant PHY0551164) for its hospitality. The valuable
spin glass server at the University of K\"oln was used to test our algorithm.
We thank M.~J\"unger and S.~Coppersmith each for a discussion of toroidal
boundary conditions.

\end{document}